\documentclass[sigconf]{acmart}

\usepackage{dirtytalk}

\copyrightyear{2021}
\acmYear{2021}
\setcopyright{acmlicensed}
\acmConference[ITiCSE 2021] {26th ACM Conference on Innovation and Technology in Computer Science Education V. 1}{June 26--July 1, 2021}{Virtual Event, Germany}
\acmBooktitle{26th ACM Conference on Innovation and Technology in Computer Science Education V. 1 (ITiCSE 2021), June 26--July 1, 2021, Virtual Event, Germany}
\acmPrice{15.00}
\acmDOI{10.1145/3430665.3456322}
\acmISBN{978-1-4503-8214-4/21/06}

\acmSubmissionID{ifp091}

\begin{document}

\fancyhead{}

\title{Students Struggle to Explain Their Own Program Code}

\newcommand{\AaltoUni}{Aalto University}

\author{Teemu Lehtinen}
\email{teemu.t.lehtinen@aalto.fi}
\orcid{0000-0003-4794-3818}
\affiliation{\institution{\AaltoUni}}

\author{Aleksi Lukkarinen}
\email{aleksi.lukkarinen@aalto.fi}
\orcid{0000-0002-3827-6243}
\affiliation{\institution{\AaltoUni}}

\author{Lassi Haaranen}
\email{lassi.haaranen@aalto.fi}
\orcid{0000-0002-6500-6425}
\affiliation{\institution{\AaltoUni}}

\begin{abstract}
We asked students to explain the structure and execution of their small programs after they had submitted them to a programming exercise. These questions about learner's code (QLCs) were delivered at three occasions in an online and open course in introductory programming as a part of the digital learning material. We make inductive content analysis to research the open-ended text answers we collected. One third of the students struggled to explain their own program code. This estimates possible occurrences of fragile learning at the moment when a student seemingly succeeds in a program writing exercise. Furthermore, we examine correlations between the correctness of the answers with other learning data. Our results indicate that answering properly aligned QLCs correctly has stronger correlation with student success and retention than merely submitting a correct program. Additionally, we present observations on learning event-driven programming to explore QLCs' potential in identifying students' thinking process.
\end{abstract}



\maketitle

\section{Introduction}
Multiple studies have investigated students' ability to explain given program code ``in plain English''~\cite{weeda2020towards,corney2014explain,murphy2012explain} or by answering multiple-choice questions~\cite{simon2014multiple}. Similarly, concept inventories for introductory programming typically include example programs and related questions that test understanding of the concepts utilized in the examples~\cite{caceffo2016developing}. Successfully explaining or tracing given program code (executing statements by hand and keeping record of the variable values) correlates with the student's ability of writing similar program code themselves~\cite{murphy2012ability,lopez2008relationships}.

Despite the correlation, an assumption that a student who created a correctly behaving program would automatically have a good understanding of the concepts involved is incorrect~\cite{kennedy2019qualitative,madison2002modular}. Students' process of arriving at a functionally correct program may include elements of guessing, copying, and hidden misconceptions that instructors should know about in order to design learning interventions. We present research on three cases of asking students questions about the structure and execution of the program they had submitted for automatic assessment. The questions share similarities to tasks cited above with the difference of targeting the learner's own program code instead of given example codes.

This research reports on delivering {\em Questions about Learner's Code} (QLCs)~\cite{lehtinen2021lets} in an online introductory programming course. In this analysis we expand from programming fundamentals towards a less researched area of learning event-driven programming. Our research answers to three questions:

\textbf{RQ1:} How well students answer QLCs? 

\textbf{RQ2:} How success in QLCs correlates with other learning data? 

\textbf{RQ3:} What potential QLCs have in  student thinking research? 

In \autoref{sec:related}, we discuss related work on students' programming process, fragile learning, and program comprehension. In \autoref{sec:method}, we describe our method in detail for further research evaluation and replication studies. In \autoref{sec:results}, we report results and in \autoref{sec:thinking} we discuss the students' answers. We finish with discussion on implications of the research in \autoref{sec:conclusion}.

\section{Related Work}
\label{sec:related}

In contrast to programming exercises and paper exams, a lab examination, where instructors silently surveyed students' programming process---including approach to error messages, testing, and available documentation---has been argued to provide improved assessment over the systematic ability to develop programs~\cite{bennedsen2007assessing}. Individual students' programming processes have been researched with think-aloud protocol which identified cases where students were able to produce correctly behaving programs while expressing uncertainty about the underlying concepts~\cite{kennedy2019qualitative}. Analysis of two students during a semester revealed that while they could write original programs that produced correct results they both had a fundamental misconception~\cite{madison2002modular}. Sometimes the delivered product---a program created by a student---can conceal fragile learning issues. This research employs QLCs to expand assessment and emphasis towards the programming process in online programming assignments.

Many students who completed programming courses have displayed symptoms of fragile learning, such as poor grasp of programming principles or difficulties in tracing the execution of a program~\cite{lister2004multinational}. While learning to program inherently involves developing original programs, recent research highlights program comprehension as a critical component~\cite{izu2019fostering}. Block Model offers a framework to classify program comprehension on two dimensions---from single elements towards the whole program; and simultaneously from text surface towards purpose~\cite{schulte2008block}. In the range of these dimensions, program comprehension can be rehearsed and assessed using example program code and different tasks, such as identifying, describing, or tracing a program and its elements~\cite{izu2019fostering}. Among these tasks, ``Explain in plain english'' matches how QLCs were designed in this research. Approximately one third of introductory programming students has previously failed ``Explain in plain English'' tasks targeting small example programs~\cite{murphy2012explain}. 

In contrast to previous work in program comprehension, our tasks target students' own programs. Students have previously described their own program code in collaborative tasks---such as pair programming~\cite{cao2005activity}---and when reviewing their program code with instructors. However, we could not identify previous research on questions that a student has received about their own program code or their answers to them. Considering the possibilities of QLCs~\cite{lehtinen2021lets}, this research evaluates manually crafted open-text questions that are posed after an automatically assessed program writing exercise.

Our introductory programming context introduces \textit{Event-Driven Programming} (EDP) concepts. Little experimental research on teaching and learning EDP has been presented to date \cite{Lukkarinen:EDPReview}. Based on personal opinions and anecdotal evidence, many authors have suggested that EDP increases the complexity of programming and might even be too difficult to teach for the novice programmer. A report on educational practice presents a controversial opinion about EDP---with the aid of a proper scaffolding---being simple enough for introductory programming education~\cite{Bruce:EdpIsSimpleEnoughForCS1} and that it might be beneficial from the perspective of teaching other programming-related concepts~\cite{Bruce:EdpFacilitatesLearning}. Through researching QLCs potential, we present our observations on possible student confusions about EDP.

\section{Method}
\label{sec:method}

\subsection{Education Context}

\begin{figure}[h]
    \includegraphics[width=\columnwidth]{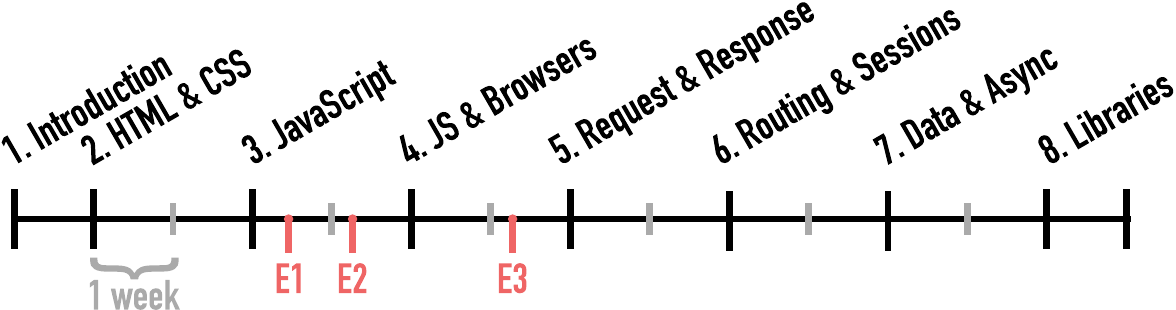}
    \caption{Structure of the course with QLCs E1, E2, and E3}
    \label{fig:iwdap}
\end{figure}

Our research was conducted in an introductory CS course in Finland, taught in English, it aims to teach the basics of programming in a web development context. The first half of the course is focused on the basics of programming and HTML \textsc{\&} CSS. The latter half is centered around HTTP communication and client-server model with Node.js. The only programming language discussed and used is JavaScript. Figure~\ref{fig:iwdap} shows the course structure and roughly where QLCs were located.

\vspace{-2mm}
\vspace{-2mm}

The intended audience for the course are people in the workforce aiming to expand their skills and enrolling is restricted to those currently not studying for a degree in a Finnish university. The broad audience also means that the backgrounds of the students are very varied. For some, this is their first time programming and some have years of industry experience in software development. The course ran for the first time in spring 2020 and the data for this research was gathered in the second instance during fall 2020. 

We had 146 students that responded to the enrolment survey, out of which 125 (86\%) gave research consent. Out of all the enrolled students, 70 of them received at least a passing grade (48\%).

\subsection{Three Exercises with QLCs}
\label{sec:exercises}

In the three exercises including QLCs (marked with E1, E2, \textsc{\&} E3 in Figure~\ref{fig:iwdap}), students first did a regular programming exercise in the online learning material where they received detailed instructions and input-output examples for a program that they must create in their IDE. The exercises also included supporting files and a possible code template for download. After students worked in their IDE they submitted their program for automated assessment. They received feedback from functional tests that were designed with a constructive agenda. Additionally, the assessment system awarded points towards completing the exercise and the course. 
Next, we present the three exercises before discussing the QLCs.

\subsubsection{Function (E1)}

The chapter introduced principles of variables and functions ending with an exercise to write a function based on a description and an input-output sample (a separate HTML page to test the student function was provided in the project):
\begin{quote}\begin{small}
The function should take two arguments. The first one should be a note name, e.g. C, and the second one is the pitch as a number, e.g. 3. Given these two arguments, the function should return them together. For example: \texttt{let note = getPitchedNote("G", 4); console.log(note); // This should print "G4"}.
\end{small}\end{quote}

\subsubsection{Loops (E2)}

The chapter containing E2 ended with an exercise that required the use of loops for printing specific lines of text repeating in a pattern. The use of two nested loop structures was enforced and guided by automated assessment. The exercise included description and an output sample we decided to omit from this paper:

\begin{quote}\begin{small}
Your task is to print out a countdown for launching multiple rockets. You need to launch 3 rockets altogether. Before starting each countdown, you need to print \texttt{Launching rocket 1} (for the first rocket) to console.log. Then you need to print a countdown from 5 to 0, and after that you need to print \texttt{Lift off!} before starting the next rocket countdown.
\end{small}\end{quote}

\subsubsection{Event-Driven Programming (E3)}
\label{sec:exercices:edp}

Round 4 (see Figure~\ref{fig:iwdap}) introduced programming in web browser context and EDP. An exercise for students aiming above passing grade included a small project template for download. Students were requested to (1)~register and complete an event handler for a set of buttons that represented two octaves of the chromatic scale as the keys of the piano as well as (2)~program event handling for buttons that start and stop recording. While recording, the program was expected to save the names and the octave numbers of the played notes to a global array. When the recording was stopped, the content of this array was to be presented to the user by creating new list items for an unordered list (\textsc{ul}) element in the Document Object Model of the same web page. Playing back the recorded notes as sound or in the correct rhythm were not part of the assignment. Screen capture of a correctly working program was presented.

\subsubsection{QLCs design}

After the previous program writing step the research cases differed from a regular exercise in that the learning material included a manually crafted form titled \say{Questions About Your Program Code.\kern-0.1em} This form included two or three QLCs that students were asked to answer using a single line text field. Apart from the single line shown for students the length of the answer was not controlled. Any answers were automatically accepted and awarded fixed points that represented 6--15\% of the exercise total and less than 0.4\% towards passing the course.

The QLCs are given in \autoref{tab:qlcs}. Generally, we aimed the QLCs to encourage reflection on the chapter content in relation to the student's work. The first QLC was designed to provide unambiguous answers that could be automatically assessed in future. The later QLCs were intentionally open-ended to allow students to elaborate more on their thinking. The QLCs about EDP focused on the less linear program flow that characterized the newly introduced concepts. In order to address a range of difficulty in program comprehension, the different QLCs were designed to reach a number of levels of the two dimensional Block Model~\cite{schulte2008block} as labeled in \autoref{tab:qlcs}.

\subsection{Analysis Decisions}
\label{sec:analysis}

Each open-ended text answer to the QLCs in each of the three exercises were examined and compared by an author who qualitatively coded the emerging themes observed in the answers. A second author applied the developed coding again to 18\% of the answers to assess inter-rater reliability. This reached 86\% overall agreement (multirater $\kappa$free 95\% CI $[0.68-0.91]$). Disagreements were resolved by improving code definitions. We describe the answers to each of the QLCs using counts of code instances. In \autoref{sec:thinking}, we examine several answers that were assigned an interesting code.

In addition to examining the answers to QLCs independently, we researched correlations with student enrolment survey, student success, and student retention. As a measure of success, we examine the ratio of awarded exercise points in four chapters following the three exercises. The time range of four chapters was an arbitrary decision that would yield seemingly continuous data and have negligible effect of student drop-out. As a measure of retention, we examine the ratio of exercises that students attempted in the last four rounds of the course. Again, the range of four rounds was an arbitrary decision based on getting adequately continuous data for statistical analysis.

In our evaluation we are interested in three different student populations for each exercise separately:

\textbf{P1:} Students submitted at least one program for the exercise but they failed to pass all functional tests in the automated assessment system. 

\textbf{P2:} Students submitted a program that did pass all of the functional tests but their answers to the related QLCs included at least one answer that was coded incorrect or irrelevant

\textbf{P3:} Students both passed all of the functional tests and correctly answered the related QLCs.

From the previous cohort on an identical course we know that distributions of both exercise points and exercise interactions experience severe ceiling and flooring effects: many students are either at the minimum or at the maximum of a selected scale. We expect this to rise partly from the exercise design---most students earn full points given adequate effort and support for an exercise---and partly from high dropout. For these distributions, where means approach ceiling, we do not want to make normality assumptions required for common t-tests. For visual evaluation, we choose to present distributions in detail using histograms. We assume that observations are independent \textsc{\&} continuous and distributions are equal apart from possible shifts in location. Then we used Wilcoxon-Mann-Whitney two-sample rank-sum test to argue any difference in medians between selected populations. In addition, results include an estimate of median difference using a 0.95 confidence interval.

\subsection{Reliability and Validity}

Initially, self-selection bias in this study is small and the participating students represent the course cohort well in exercises E1 and E2. Unfortunately, the third exercise E3 had significant self-selection bias that must be considered in analysis. Our students' varied background could affect the results and for that reason we report correlations with data from our background survey.

Above, we tested inter-rater reliability of the coding used. Analysis could focus on other topics that develop different criteria and coding than in this research. However, we are confident in detecting incorrect answers which are relevant for correlation study.

The decided success and retention measures and their arbitrary time range choices are a threat to validity. We see success as a partial course grade earned in relevant time span and retention as working until the end of the course material. However, we argue any measure of success or retention on a course are less than ideal interpretation of learning and valid interpretations should replicate the same results to a certain degree.

More importantly, students' ability to answer QLCs depends on the following factors: questions' alignment with teaching, the amount \textsc{\&} quality of teaching \textsc{\&} practice students have had, the experience students have in program comprehension tasks, the design of the programming exercises that the QLCs target, and the wording of the questions. In the interest of replication and variations of this research, we aim to report such factors in detail. This early research aims to report potential and correlations of QLCs and not to validate instruments or causalities.

\section{Results}
\label{sec:results}
 
 \subsection{Student Answers to QLCs (RQ1)}

\begin{table}
\caption{Answer coding and frequencies for QLCs by students' program passing or failing related functional tests. {\small Targeted Block Model levels are given in square brackets.}}
\label{tab:qlcs}
\centering
\begin{tabular}{p{0.65\linewidth}rrr}
\toprule
\multicolumn{4}{p{0.95\linewidth}}{\raggedright\textbf{E1.1} Which are the two function parameter names in your program? {\it\small [BM: atom -- text surface]}} \\
& {\it pass} & {\it fail} & $N=82$ \\
- Incorrect or irrelevant answer &   1 &   3 &   5\% \\ 
+ Correct parameter names &  70 &   8 &  95\% \\
\midrule
\multicolumn{4}{p{0.95\linewidth}}{\raggedright\textbf{E1.2} Where would you guess the parameter values come from when you test in the provided browser page synth.html? \mbox{{\it\small [BM: relation -- text surface]}}} \\
& {\it pass} & {\it fail} & $N=82$ \\
- Incorrect or irrelevant answer & 19 & 8 & 33\% \\
+ The function call in source &  15 &   1 &  19\% \\
+ The argument variables traced backwards &  30 &   2 &  39\% \\
+ The user task generating values &   7 &   0 &   9\% \\
\midrule
\multicolumn{4}{p{0.95\linewidth}}{\raggedright\textbf{E1.3} How would you describe the difference between returning a~value and printing a~value using console.log? \mbox{{\it\small [BM: atom -- execution]}}} \\
& {\it pass} & {\it fail} & $N=82$ \\
- Incorrect or irrelevant answer & 16 & 7 & 28\% \\
+ Returning a value (vague) &  16 &   1 &  21\% \\ 
+ Returning a value for further computation &  32 &   2 &  41\% \\ 
+ Logging a value to view/debug it &  59 &   4 &  77\% \\ 
\midrule
\multicolumn{4}{p{0.95\linewidth}}{\raggedright\textbf{E2.1} Describe the responsibilities of your outer loop in few words. {\it\small [BM: block -- execution]}} \\
& {\it pass} & {\it fail} & $N=68$ \\
- Incorrect or irrelevant answer & 13 & 1 & 20\% \\
+ Iterates/prints rockets (incomplete) &  34 &   2 &  53\% \\ 
+ Iterates rockets \textsc{\&} commences countdown &  18 &   0 &  27\% \\ 
\midrule
\multicolumn{4}{p{0.95\linewidth}}{\raggedright\textbf{E2.2} Describe the responsibilities of your inner loop in few words. {\it\small [BM: block -- execution]}} \\
& {\it pass} & {\it fail} & $N=68$ \\
- Incorrect or irrelevant answer & 15 & 2 & 25\% \\
+ Prints countdown (no extraneous parts) &  50 &   1 &  75\% \\
\midrule
\multicolumn{4}{p{0.95\linewidth}}{\raggedright\textbf{E3.1} Which parts of your program execute when browser opens the page? {\it\small [BM: macro -- execution]}} \\
& {\it pass} & {\it fail} & $N=33$ \\
- Incorrect or irrelevant answer & 10 & 4 & 42\% \\
+ Declarations &  11 &   1 &  36\% \\ 
+ Event listeners (vague) &   5 &   1 &   18\% \\ 
+ Registering event listeners &  10 &   1 &  33\% \\ 
\midrule
\multicolumn{4}{p{0.95\linewidth}}{\raggedright\textbf{E3.2} Which parts of your program execute when the user clicks one of the note buttons? {\it\small [BM: macro -- execution]}} \\
& {\it pass} & {\it fail} & $N=33$ \\
- Incorrect or irrelevant answer & 5 & 2 & 21\% \\
+ Event (vague) &   4 &   2 &   18\% \\ 
+ Event listener &  18 &   2 &  61\% \\ 
\midrule
\multicolumn{4}{p{0.95\linewidth}}{\raggedright\textbf{E3.3} Describe in a few words how your program remembers the played notes. {\it\small [BM: macro -- purpose]}} \\
& {\it pass} & {\it fail} & $N=33$ \\
- Incorrect or irrelevant answer & 2 & 1 & 9\% \\
+ Stored in an array variable &  26 &   5 &  91\% \\ 
\bottomrule
\end{tabular}
\end{table}

\begin{figure}
    \includegraphics{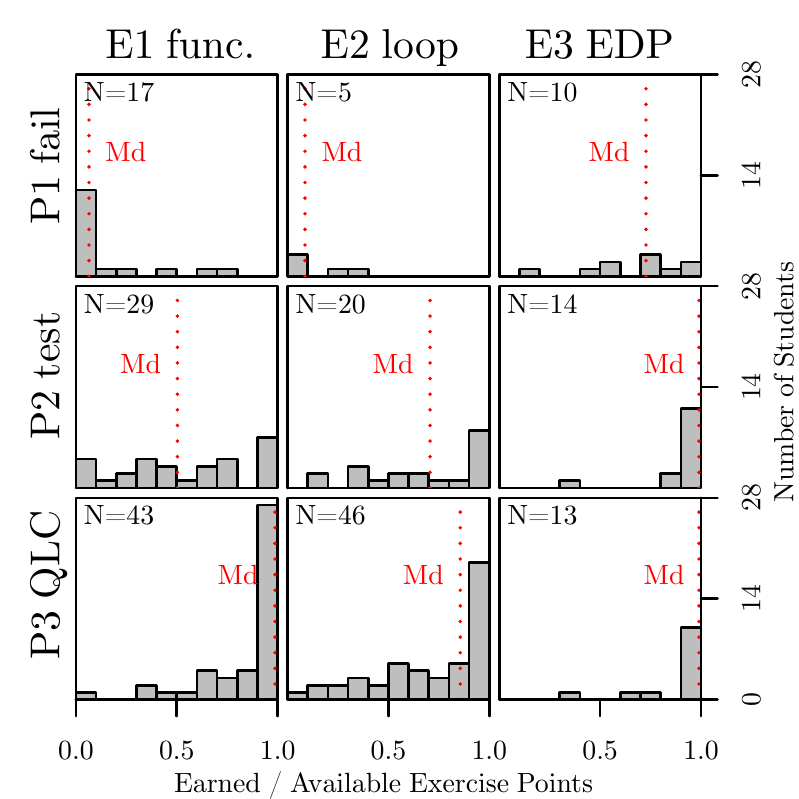}
    \caption{Frequencies of \emph{exercise success} following three separate exercises and three different student populations for them: \small{P1 submitted programs failing functional tests, P2 passed tests but answered QLCs incorrectly, and P3 both passed tests and answered QLCs correctly.}}
    \label{fig:success}
\end{figure}

\begin{figure}
    \includegraphics{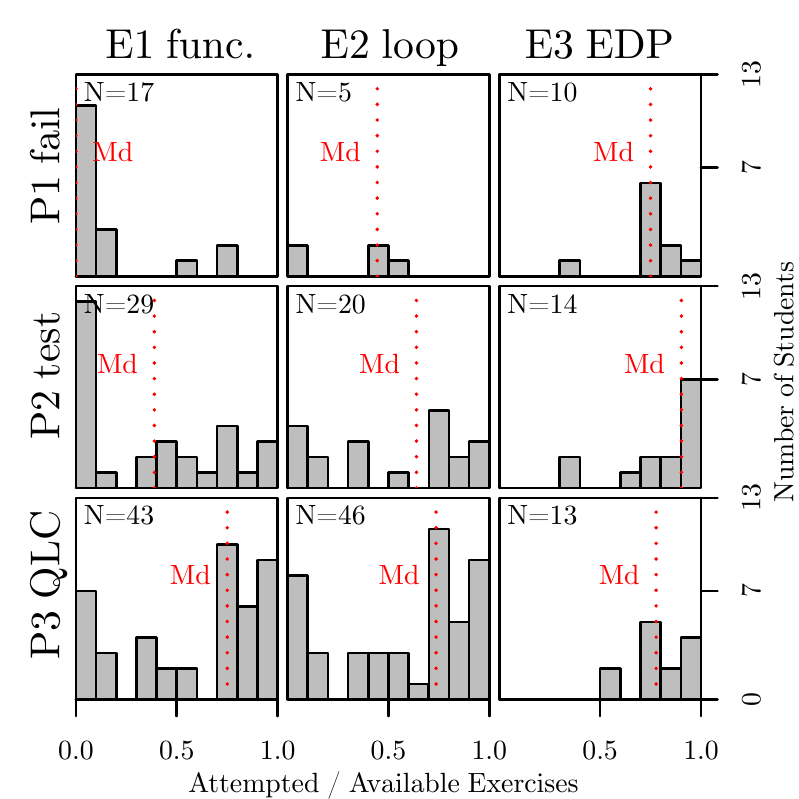}
    \caption{Frequencies of \emph{attempted exercises} towards end of the course divided by three separate exercises and three different student populations for them: \small{P1 submitted programs failing functional tests, P2 passed tests but answered QLCs incorrectly, and P3 both passed tests and answered QLCs correctly.}}
    \label{fig:retention}
\end{figure}

\autoref{tab:qlcs} presents the answer coding and code frequencies. For E1 approximately one third of the students failed to answer at least one of the three questions correctly. Four incorrect answers for parameter names in E1.1 included \textit{\say{C and 3}} and \textit{\say{Absolutely no idea.}} Common incorrect answers for E1.2 were file names that did not call the function in the given template. For E1.3 incorrect answers were longer more unique descriptions, such as \textit{\say{returning a value is like running it, console.log is the test version of it.}}

Little less than one third of the students failed to answer either E2.1 or E2.2 correctly. Many of the answers were rather irrelevant discussions, such as \textit{\say{Outer loop executes first.}} A repeating mistake was to claim inner loop prints \say{Lift off!} while it clearly is outside the countdown.

The QLCs about EDP in E3 were the hardest to answer. Approximately half of the students failed at least one of the questions and many correct answers were vague. Importantly, many students did not reach E3 at all in time or decided to skip it. Students described this chapter as demanding and laborious. Our research in E3 should be seen more as exploratory work, which we discuss in \autoref{sec:thinking}.

\subsection{Correlations With Answers to QLCs (RQ2)}

Answering correctly to QLCs had no correlation with age, gender, level of education, self-reported determination of success on course, or students considering themselves as beginners in programming. Student's earlier participation in programming courses was moderately correlated with answering correctly, $r(109)=.43, p<.001$. There was no correlation with the education level of prior programming courses. Additionally, the student's estimate of prior programming hours was moderately correlated with answering correctly to QLCs, $r(102)=.37, p<.001$.

\autoref{fig:success} presents student success after each of the three exercises as three frequency histograms for different student populations.
For the exercises about function (E1) and loops (E2) it is evident that a small number of students who failed the functional tests (P1) earned least points in the following four chapters. Furthermore, students who both passed the tests and answered QLCs correctly (P3) earned more points than students who answered QLCs incorrectly (P2). For E1 medians for P2 and P3 are $0.50$ and $0.98$ having difference 95\% CI $[-0.48,-0.07]$ which is statistically significant (Mann-Whitney $U=299, n_1=27, n_2=45, P<.001$ two-tailed). For E2 medians for P2 and P3 are $0.71$ and $0.86$ having difference 95\% CI $[-0.20,0.08]$ which is \emph{insignificant} (Mann-Whitney $U=410, n_1=20, n_2=46, P=.483$ two-tailed). Regarding the exercise about EDP (E3), we believe the low number of students and most observations at ceiling prevented us from making any arguments or difference tests.

\autoref{fig:retention} presents student retention depending on the three exercises that each divide students into three populations presented as frequency histograms. Among the students that failed the functional tests (P1) in the exercises about function (E1) and loops (E2), nobody attempted all of the exercises in the last four rounds. Additionally, students who both passed the tests and answered QLCs correctly (P3) attempted more exercises than students who answered QLCs incorrectly (P2). For E1, medians for P2 and P3 are $0.33$ and $0.75$ having difference 95\% CI $[-0.42,-0.03]$ which is statistically significant (Mann-Whitney $U=382, n_1=27, n_2=45, P=.008$ two-tailed). For E2, medians for P2 and P3 are $0.64$ and $0.74$ having difference 95\% CI $[-0.25,0.14]$ which is \emph{insignificant} (Mann-Whitney $U=413, n_1=20, n_2=46, P=.514$ two-tailed). The exercise about EDP (E3) breaks the pattern and students who passed the tests but answered QLCs incorrectly (P2) attempted most exercises. All of the students who at least attempted E3 worked a good amount in the last rounds of the course.

\section{Student Thinking (RQ3)}
\label{sec:thinking}

In this chapter, we discuss potential reasons underlying the answers. We examine student thinking through our answer coding and selected answer quotations. The open feedback from the course suggests that QLCs do make students reflect. For instance, two students wrote about \textit{\say{They can kind of check if the student completed the task him/herself and understood what was done and why.}} and 

\begin{quote}\begin{small}
\textit{\say{I don't know whether these questions helped me. After all, often they asked for things one already thinks about when writing the code. So it felt a bit redundant at times. But then again, maybe it helps with self-reflection.}}
\end{small}\end{quote}

\subsection{Answering Challenges}

The QLCs in this research require a varying levels of program comprehension to answer correctly. By examining the incorrect answers, we identified three challenges that manifested in multiple QLCs.

\subsubsection{Defective Expression}
Many of the answers evoked uncertainty of whether the root cause for an incorrect answer is the respondent's understanding or merely a~defective written expression. While English was used for both presenting the questions and giving the answers, many of the respondents are not native English speakers. Consequently, an inadequate command of English might pose challenges in expressing one's thoughts precisely. In addition to possible challenges in using English, possible earlier issues with clear understanding of prerequisite concepts might result in non-ideal word choices.

One example of a defective expression can be found in an answer to E1.3:
\textit{\say{Returned value can be used elsewhere in the program, printed value can't be accessed.}} Certainly a value can be accessed after printing it. However, if a return statement is replaced with a print statement, the value becomes inaccessible from outside the function scope. Another example from E3.1 is the usage of the verb \textit{to activate} in \textit{\say{Event listeners are activated.}} We can ask if it means that the event handlers are being attached (registered to be called later) or executed (actually called to handle an event).

\subsubsection{Misunderstood Terminology} The course teaches new terminology, which confused some of the students. There might also be clashes between synonymous terms that are being used in different contexts (e.g., development environments, teaching materials, discussions with people) or interchangeably in one context. For instance, the answer \textit{\say{Eventlisteners and then eventhandlers}} (E3.2) illustrates that the respondent most likely has a confusion with the synonymous terms \textit{event listener} and \textit{event handler,} both of which mean a subprogram that is called to respond to an event. Unless this respondent has some special reason for calling event handlers with two synonymous terms, they might have associated some differing meaning for one of these terms.

\subsubsection{Concept vs. Application} Some answers contained terms written in camel case or without spaces, as identifiers are written in some programming languages. The respondent might have written them themselves, thinking for some reason that they have to be written that way, or they might have copied those terms from some program code. One hypothesis for the reason of this behavior is that the respondent is following a surface learning approach, focusing on passing the exercises quickly without thinking on conceptual level what the code actually contains or does. This would result, for instance, in copying a name of an event handler (such as \textit{\say{eventHandler}}) from their code to the answer instead of using the name of the concept. A similar situation could exist with an answer \textit{\say{variable definitions and addeventlisteners.}} There, the (presumed) method name \textit{\say{addeventlisteners}} is being used instead of describing the attachment of the event handlers on the conceptual level, as was done with variables in the same answer.

\subsection{EDP-Related Confusions}

From students' answers to the QLCs, we identified possible confusions about EDP and, more specifically, program flow, which was at focus in E3. As these observations are based on the answers of a few individual students,we cannot predict the true frequencies of the presented confusions among the whole student base.

\subsubsection{Loading-Time Behavior} The answers of six respondents (18\%) to E3.1 indicated that static resources, such as the HTML and CSS files, would be the program parts executed when a web page is opened in a browser. One hypothesis for this would be that the respondents have an elementary-level confusion about what resources are static and what are executable program code. However, this seems somewhat implausible, as the participants have been writing JavaScript on the course already, and one of these respondents explicitly used the word \textit{static} (\textit{\say{It goes over the static html file}}).

Another reason for referring to static resources could be that instead of understanding \textit{a program} to refer to the JavaScript code they are editing, these respondents reason that their program is composed of all the files in the exercise and that the browser first loads the HTML and CSS files. A third possibility is that the respondents have an unclear notion of \textit{executing a program}, and consequently they think that executing means, for instance, loading and parsing all the related files and then displaying the resulting web page.

\subsubsection{Attaching and Executing Event Handlers.} Seven responses (21\%) for E3.1 and E3.2 implied that some of the respondents might be confused about what it means to attach (to add, to register) an event handler, when they are actually executed, and who actually calls them. In other words, details of essential event handling mechanics, including the \textit{Observer} design pattern, potentially pose confusions at least when left explicitly unexplained.

Concerning E3.1, the responses \textit{\say{event listeners are activated,}} \textit{\say{all the global variables and event-listeners,}} and \textit{\say{eventListeners}} could indicate the respondents to think that the event handlers are executed at the time of attaching them---in this case, when the web page is loaded. Moreover, E3.2 gathered such explanations as \textit{\say{eachNoteButtons[i].addEventListener('click', playNote) executes the playNote event handler,}} which quite clearly reflect the incorrect idea of \texttt{addEventListener()} method calling the given event handler after the occurrences of the events.

\section{Conclusion}
\label{sec:conclusion}

Our results support earlier findings~\cite{kennedy2019qualitative, madison2002modular} on students struggling with concepts they applied when writing a program. In our experiment, a similar ratio of students failed at correctly explaining their own program code as has previously been reported~\cite{murphy2012explain} for explaining example codes (RQ1). Our analysis of student thinking suggests other answering challenges besides program comprehension. We note that open-text explanations do not only require program comprehension but also the ability to describe the programming concepts adequately. We also argue that both the program comprehension and the ability to describe the program are necessary skills in a programming career.

The QLCs in E1 targeted syntax and execution of individual language elements related to functions. They were well aligned with the current learning goals and relatively simple according to the Block Model~\cite{schulte2008block}.

After submitting a functionally correct program to E1, the students that also answered correctly to the posed QLCs had significantly higher correlation with success and retention on the course (RQ2). The same phenomena was statistically insignificant for QLCs in E2. These QLCs targeted execution of code blocks---more demanding according to the Block Model---and were considerably more open-ended. We hypothesize that the open-endedness combined with the identified challenges in student answers are important factors for observing weaker correlation effects.

Finally, we analysed student thinking from the QLC answers and identified several challenges that the students had. Specially, E3 had a low number of students for statistical analysis but enabled exploratory work in the EDP context. We observed several students had confusions with loading-time behaviour as well as attaching vs. executing event handlers. Because a limited student population completed E3, the true frequency of these confusions about EDP is unknown. However, as we strive to improve our teaching in every possible way, these observations should be useful in considering areas that might potentially have been left unclear or misunderstood. We recommend that educators consider them in everyday teaching of EDP outside formal research contexts. Given that these observations were possible, we see that open-ended QLCs can be useful in researching student thinking (RQ3).

We still lack knowledge whether QLCs, as employed in this research, can be used as instruments for measuring program comprehension, knowledge of terminology, or only careful answering---and whether answering itself has learning gains via reflection or other processes. Our research and discussion support all of these hypotheses. We note that the results are also greatly dependent on the type and design of the QLCs and their application in the course. Given the QLCs' correlation with student success and retention in online learning context, we see that replications and improved research designs are worth considerable future efforts.

In our future work, we will
\begin{enumerate}
    \item develop automatic personalized QLC generation\\from the programs that students submit,
    \item evaluate QLCs' potential effect on learning, and
    \item investigate students' understanding of EDP-related\\concepts and program code.
\end{enumerate}

To drill deeper and to see whether our observations can be repeated, we encourage the computing education community to carry out replications or variations of this study in varying teaching contexts (e.g., student populations and learning materials). Regarding students' understanding of EDP, it might be helpful to complement QLCs with other research methods, including concept maps.

\bibliographystyle{ACM-Reference-Format}
\balance
\bibliography{main}


\begin{thebibliography}{18}


\ifx \showCODEN    \undefined \def \showCODEN     #1{\unskip}     \fi
\ifx \showDOI      \undefined \def \showDOI       #1{#1}\fi
\ifx \showISBNx    \undefined \def \showISBNx     #1{\unskip}     \fi
\ifx \showISBNxiii \undefined \def \showISBNxiii  #1{\unskip}     \fi
\ifx \showISSN     \undefined \def \showISSN      #1{\unskip}     \fi
\ifx \showLCCN     \undefined \def \showLCCN      #1{\unskip}     \fi
\ifx \shownote     \undefined \def \shownote      #1{#1}          \fi
\ifx \showarticletitle \undefined \def \showarticletitle #1{#1}   \fi
\ifx \showURL      \undefined \def \showURL       {\relax}        \fi
\providecommand\bibfield[2]{#2}
\providecommand\bibinfo[2]{#2}
\providecommand\natexlab[1]{#1}
\providecommand\showeprint[2][]{arXiv:#2}

\bibitem[\protect\citeauthoryear{Bennedsen and Caspersen}{Bennedsen and
  Caspersen}{2007}]%
        {bennedsen2007assessing}
\bibfield{author}{\bibinfo{person}{Jens Bennedsen} {and}
  \bibinfo{person}{Michael~E. Caspersen}.} \bibinfo{year}{2007}\natexlab{}.
\newblock \showarticletitle{Assessing Process and Product}.
\newblock \bibinfo{journal}{\emph{Innovation in Teaching and Learning in
  Information and Computer Sciences}} \bibinfo{volume}{6}, \bibinfo{number}{4}
  (\bibinfo{year}{2007}), \bibinfo{pages}{183--202}.
\newblock
\urldef\tempurl%
\url{https://doi.org/10.11120/ital.2007.06040183}
\showDOI{\tempurl}


\bibitem[\protect\citeauthoryear{Bruce and Danyluk}{Bruce and Danyluk}{2004}]%
        {Bruce:EdpFacilitatesLearning}
\bibfield{author}{\bibinfo{person}{Kim~B. Bruce} {and} \bibinfo{person}{Andrea
  Danyluk}.} \bibinfo{year}{2004}\natexlab{}.
\newblock \showarticletitle{Event-Driven Programming Facilitates Learning
  Standard Programming Concepts}. In \bibinfo{booktitle}{\emph{Companion to the
  19th Annual ACM SIGPLAN Conference on Object-Oriented Programming Systems,
  Languages, and Applications}} (Vancouver, BC, CANADA)
  \emph{(\bibinfo{series}{OOPSLA '04})}. \bibinfo{publisher}{Association for
  Computing Machinery}, \bibinfo{address}{New York, NY, USA},
  \bibinfo{pages}{96–100}.
\newblock
\showISBNx{1581138334}
\urldef\tempurl%
\url{https://doi.org/10.1145/1028664.1028704}
\showDOI{\tempurl}


\bibitem[\protect\citeauthoryear{Bruce, Danyluk, and Murtagh}{Bruce
  et~al\mbox{.}}{2001}]%
        {Bruce:EdpIsSimpleEnoughForCS1}
\bibfield{author}{\bibinfo{person}{Kim~B. Bruce}, \bibinfo{person}{Andrea~P.
  Danyluk}, {and} \bibinfo{person}{Thomas~P. Murtagh}.}
  \bibinfo{year}{2001}\natexlab{}.
\newblock \showarticletitle{Event-Driven Programming is Simple Enough for CS1}.
  In \bibinfo{booktitle}{\emph{Proceedings of the 6th Annual Conference on
  Innovation and Technology in Computer Science Education}} (Canterbury, United
  Kingdom) \emph{(\bibinfo{series}{ITiCSE '01})}.
  \bibinfo{publisher}{Association for Computing Machinery},
  \bibinfo{address}{New York, NY, USA}, \bibinfo{pages}{1–4}.
\newblock
\showISBNx{1581133308}
\urldef\tempurl%
\url{https://doi.org/10.1145/377435.377440}
\showDOI{\tempurl}


\bibitem[\protect\citeauthoryear{Caceffo, Wolfman, Booth, and Azevedo}{Caceffo
  et~al\mbox{.}}{2016}]%
        {caceffo2016developing}
\bibfield{author}{\bibinfo{person}{Ricardo Caceffo}, \bibinfo{person}{Steve
  Wolfman}, \bibinfo{person}{Kellogg~S. Booth}, {and} \bibinfo{person}{Rodolfo
  Azevedo}.} \bibinfo{year}{2016}\natexlab{}.
\newblock \showarticletitle{Developing a Computer Science Concept Inventory for
  Introductory Programming}. In \bibinfo{booktitle}{\emph{Proceedings of the
  47th ACM Technical Symposium on Computing Science Education}} (Memphis,
  Tennessee, USA) \emph{(\bibinfo{series}{SIGCSE '16})}.
  \bibinfo{publisher}{Association for Computing Machinery},
  \bibinfo{address}{New York, NY, USA}, \bibinfo{pages}{364–369}.
\newblock
\showISBNx{9781450336857}
\urldef\tempurl%
\url{https://doi.org/10.1145/2839509.2844559}
\showDOI{\tempurl}


\bibitem[\protect\citeauthoryear{Corney, Fitzgerald, Hanks, Lister, McCauley,
  and Murphy}{Corney et~al\mbox{.}}{2014}]%
        {corney2014explain}
\bibfield{author}{\bibinfo{person}{Malcolm Corney}, \bibinfo{person}{Sue
  Fitzgerald}, \bibinfo{person}{Brian Hanks}, \bibinfo{person}{Raymond Lister},
  \bibinfo{person}{Renee McCauley}, {and} \bibinfo{person}{Laurie Murphy}.}
  \bibinfo{year}{2014}\natexlab{}.
\newblock \showarticletitle{'explain in Plain English' Questions Revisited:
  Data Structures Problems}. In \bibinfo{booktitle}{\emph{Proceedings of the
  45th ACM Technical Symposium on Computer Science Education}} (Atlanta,
  Georgia, USA) \emph{(\bibinfo{series}{SIGCSE '14})}.
  \bibinfo{publisher}{Association for Computing Machinery},
  \bibinfo{address}{New York, NY, USA}, \bibinfo{pages}{591–596}.
\newblock
\showISBNx{9781450326056}
\urldef\tempurl%
\url{https://doi.org/10.1145/2538862.2538911}
\showDOI{\tempurl}


\bibitem[\protect\citeauthoryear{Izu, Schulte, Aggarwal, Cutts, Duran, Gutica,
  Heinemann, Kraemer, Lonati, Mirolo, and Weeda}{Izu et~al\mbox{.}}{2019}]%
        {izu2019fostering}
\bibfield{author}{\bibinfo{person}{Cruz Izu}, \bibinfo{person}{Carsten
  Schulte}, \bibinfo{person}{Ashish Aggarwal}, \bibinfo{person}{Quintin Cutts},
  \bibinfo{person}{Rodrigo Duran}, \bibinfo{person}{Mirela Gutica},
  \bibinfo{person}{Birte Heinemann}, \bibinfo{person}{Eileen Kraemer},
  \bibinfo{person}{Violetta Lonati}, \bibinfo{person}{Claudio Mirolo}, {and}
  \bibinfo{person}{Renske Weeda}.} \bibinfo{year}{2019}\natexlab{}.
\newblock \showarticletitle{Fostering Program Comprehension in Novice
  Programmers - Learning Activities and Learning Trajectories}. In
  \bibinfo{booktitle}{\emph{Proceedings of the Working Group Reports on
  Innovation and Technology in Computer Science Education}} (Aberdeen, Scotland
  Uk) \emph{(\bibinfo{series}{ITiCSE-WGR '19})}.
  \bibinfo{publisher}{Association for Computing Machinery},
  \bibinfo{address}{New York, NY, USA}, \bibinfo{pages}{27–52}.
\newblock
\showISBNx{9781450375672}
\urldef\tempurl%
\url{https://doi.org/10.1145/3344429.3372501}
\showDOI{\tempurl}


\bibitem[\protect\citeauthoryear{Kennedy and Kraemer}{Kennedy and
  Kraemer}{2019}]%
        {kennedy2019qualitative}
\bibfield{author}{\bibinfo{person}{Cazembe Kennedy} {and}
  \bibinfo{person}{Eileen~T. Kraemer}.} \bibinfo{year}{2019}\natexlab{}.
\newblock \showarticletitle{Qualitative Observations of Student Reasoning:
  Coding in the Wild}. In \bibinfo{booktitle}{\emph{Proceedings of the 2019 ACM
  Conference on Innovation and Technology in Computer Science Education}}
  (Aberdeen, Scotland Uk) \emph{(\bibinfo{series}{ITiCSE '19})}.
  \bibinfo{publisher}{Association for Computing Machinery},
  \bibinfo{address}{New York, NY, USA}, \bibinfo{pages}{224–230}.
\newblock
\showISBNx{9781450368957}
\urldef\tempurl%
\url{https://doi.org/10.1145/3304221.3319751}
\showDOI{\tempurl}


\bibitem[\protect\citeauthoryear{{Lan Cao} and {Peng Xu}}{{Lan Cao} and {Peng
  Xu}}{2005}]%
        {cao2005activity}
\bibfield{author}{\bibinfo{person}{{Lan Cao}} {and} \bibinfo{person}{{Peng
  Xu}}.} \bibinfo{year}{2005}\natexlab{}.
\newblock \showarticletitle{Activity Patterns of Pair Programming}. In
  \bibinfo{booktitle}{\emph{Proceedings of the 38th Annual Hawaii International
  Conference on System Sciences}}. \bibinfo{publisher}{IEEE},
  \bibinfo{address}{New York, NY, USA}, \bibinfo{pages}{88a--88a}.
\newblock
\urldef\tempurl%
\url{https://doi.org/10.1109/HICSS.2005.66}
\showDOI{\tempurl}


\bibitem[\protect\citeauthoryear{Lehtinen, Santos, and Sorva}{Lehtinen
  et~al\mbox{.}}{pear}]%
        {lehtinen2021lets}
\bibfield{author}{\bibinfo{person}{Teemu Lehtinen}, \bibinfo{person}{Andr\'e~L.
  Santos}, {and} \bibinfo{person}{Juha Sorva}.} \bibinfo{year}{to
  appear}\natexlab{}.
\newblock \showarticletitle{Let's Ask Students About Their Programs,
  Automatically}. In \bibinfo{booktitle}{\emph{Proceedings of the 28th
  International Conference on Program Comprehension}}
  \emph{(\bibinfo{series}{ICPC '21})}. \bibinfo{publisher}{IEEE},
  \bibinfo{address}{New York, NY,USA}.
\newblock


\bibitem[\protect\citeauthoryear{Lister, Adams, Fitzgerald, Fone, Hamer,
  Lindholm, McCartney, Mostr\"{o}m, Sanders, Sepp\"{a}l\"{a}, Simon, and
  Thomas}{Lister et~al\mbox{.}}{2004}]%
        {lister2004multinational}
\bibfield{author}{\bibinfo{person}{Raymond Lister},
  \bibinfo{person}{Elizabeth~S. Adams}, \bibinfo{person}{Sue Fitzgerald},
  \bibinfo{person}{William Fone}, \bibinfo{person}{John Hamer},
  \bibinfo{person}{Morten Lindholm}, \bibinfo{person}{Robert McCartney},
  \bibinfo{person}{Jan~Erik Mostr\"{o}m}, \bibinfo{person}{Kate Sanders},
  \bibinfo{person}{Otto Sepp\"{a}l\"{a}}, \bibinfo{person}{Beth Simon}, {and}
  \bibinfo{person}{Lynda Thomas}.} \bibinfo{year}{2004}\natexlab{}.
\newblock \showarticletitle{A Multi-National Study of Reading and Tracing
  Skills in Novice Programmers}. In \bibinfo{booktitle}{\emph{Working Group
  Reports from ITiCSE on Innovation and Technology in Computer Science
  Education}} (Leeds, United Kingdom) \emph{(\bibinfo{series}{ITiCSE-WGR
  '04})}. \bibinfo{publisher}{Association for Computing Machinery},
  \bibinfo{address}{New York, NY, USA}, \bibinfo{pages}{119–150}.
\newblock
\showISBNx{9781450377942}
\urldef\tempurl%
\url{https://doi.org/10.1145/1044550.1041673}
\showDOI{\tempurl}


\bibitem[\protect\citeauthoryear{Lopez, Whalley, Robbins, and Lister}{Lopez
  et~al\mbox{.}}{2008}]%
        {lopez2008relationships}
\bibfield{author}{\bibinfo{person}{Mike Lopez}, \bibinfo{person}{Jacqueline
  Whalley}, \bibinfo{person}{Phil Robbins}, {and} \bibinfo{person}{Raymond
  Lister}.} \bibinfo{year}{2008}\natexlab{}.
\newblock \showarticletitle{Relationships between Reading, Tracing and Writing
  Skills in Introductory Programming}. In \bibinfo{booktitle}{\emph{Proceedings
  of the Fourth International Workshop on Computing Education Research}}
  (Sydney, Australia) \emph{(\bibinfo{series}{ICER '08})}.
  \bibinfo{publisher}{Association for Computing Machinery},
  \bibinfo{address}{New York, NY, USA}, \bibinfo{pages}{101–112}.
\newblock
\showISBNx{9781605582160}
\urldef\tempurl%
\url{https://doi.org/10.1145/1404520.1404531}
\showDOI{\tempurl}


\bibitem[\protect\citeauthoryear{Lukkarinen, Malmi, and Haaranen}{Lukkarinen
  et~al\mbox{.}}{2021}]%
        {Lukkarinen:EDPReview}
\bibfield{author}{\bibinfo{person}{Aleksi Lukkarinen}, \bibinfo{person}{Lauri
  Malmi}, {and} \bibinfo{person}{Lassi Haaranen}.}
  \bibinfo{year}{2021}\natexlab{}.
\newblock \showarticletitle{Event-Driven Programming in Programming Education:
  A Mapping Review}.
\newblock \bibinfo{journal}{\emph{ACM Transactions on Computing Education
  (TOCE)}} \bibinfo{volume}{21}, \bibinfo{number}{1}, Article
  \bibinfo{articleno}{1} (\bibinfo{date}{March} \bibinfo{year}{2021}),
  \bibinfo{numpages}{31}~pages.
\newblock
\urldef\tempurl%
\url{https://doi.org/10.1145/3423956}
\showDOI{\tempurl}


\bibitem[\protect\citeauthoryear{Madison and Gifford}{Madison and
  Gifford}{2002}]%
        {madison2002modular}
\bibfield{author}{\bibinfo{person}{Sandra Madison} {and} \bibinfo{person}{James
  Gifford}.} \bibinfo{year}{2002}\natexlab{}.
\newblock \showarticletitle{Modular Programming}.
\newblock \bibinfo{journal}{\emph{Journal of Research on Technology in
  Education}} \bibinfo{volume}{34}, \bibinfo{number}{3} (\bibinfo{year}{2002}),
  \bibinfo{pages}{217--229}.
\newblock
\urldef\tempurl%
\url{https://doi.org/10.1080/15391523.2002.10782346}
\showDOI{\tempurl}


\bibitem[\protect\citeauthoryear{Murphy, Fitzgerald, Lister, and
  McCauley}{Murphy et~al\mbox{.}}{2012a}]%
        {murphy2012ability}
\bibfield{author}{\bibinfo{person}{Laurie Murphy}, \bibinfo{person}{Sue
  Fitzgerald}, \bibinfo{person}{Raymond Lister}, {and}
  \bibinfo{person}{Ren\'{e}e McCauley}.} \bibinfo{year}{2012}\natexlab{a}.
\newblock \showarticletitle{Ability to 'explain in Plain English' Linked to
  Proficiency in Computer-Based Programming}. In
  \bibinfo{booktitle}{\emph{Proceedings of the Ninth Annual International
  Conference on International Computing Education Research}} (Auckland, New
  Zealand) \emph{(\bibinfo{series}{ICER '12})}. \bibinfo{publisher}{Association
  for Computing Machinery}, \bibinfo{address}{New York, NY, USA},
  \bibinfo{pages}{111–118}.
\newblock
\showISBNx{9781450316040}
\urldef\tempurl%
\url{https://doi.org/10.1145/2361276.2361299}
\showDOI{\tempurl}


\bibitem[\protect\citeauthoryear{Murphy, McCauley, and Fitzgerald}{Murphy
  et~al\mbox{.}}{2012b}]%
        {murphy2012explain}
\bibfield{author}{\bibinfo{person}{Laurie Murphy}, \bibinfo{person}{Ren\'{e}e
  McCauley}, {and} \bibinfo{person}{Sue Fitzgerald}.}
  \bibinfo{year}{2012}\natexlab{b}.
\newblock \showarticletitle{'Explain in Plain English' Questions: Implications
  for Teaching}. In \bibinfo{booktitle}{\emph{Proceedings of the 43rd ACM
  Technical Symposium on Computer Science Education}} (Raleigh, North Carolina,
  USA) \emph{(\bibinfo{series}{SIGCSE '12})}. \bibinfo{publisher}{Association
  for Computing Machinery}, \bibinfo{address}{New York, NY, USA},
  \bibinfo{pages}{385–390}.
\newblock
\showISBNx{9781450310987}
\urldef\tempurl%
\url{https://doi.org/10.1145/2157136.2157249}
\showDOI{\tempurl}


\bibitem[\protect\citeauthoryear{Schulte}{Schulte}{2008}]%
        {schulte2008block}
\bibfield{author}{\bibinfo{person}{Carsten Schulte}.}
  \bibinfo{year}{2008}\natexlab{}.
\newblock \showarticletitle{Block Model: An Educational Model of Program
  Comprehension as a Tool for a Scholarly Approach to Teaching}. In
  \bibinfo{booktitle}{\emph{Proceedings of the Fourth International Workshop on
  Computing Education Research}} (Sydney, Australia)
  \emph{(\bibinfo{series}{ICER '08})}. \bibinfo{publisher}{Association for
  Computing Machinery}, \bibinfo{address}{New York, NY, USA},
  \bibinfo{pages}{149–160}.
\newblock
\showISBNx{9781605582160}
\urldef\tempurl%
\url{https://doi.org/10.1145/1404520.1404535}
\showDOI{\tempurl}


\bibitem[\protect\citeauthoryear{Simon and Snowdon}{Simon and Snowdon}{2014}]%
        {simon2014multiple}
\bibfield{author}{\bibinfo{person}{Simon} {and} \bibinfo{person}{Susan
  Snowdon}.} \bibinfo{year}{2014}\natexlab{}.
\newblock \showarticletitle{Multiple-Choice vs Free-Text Code-Explaining
  Examination Questions}. In \bibinfo{booktitle}{\emph{Proceedings of the 14th
  Koli Calling International Conference on Computing Education Research}}
  (Koli, Finland) \emph{(\bibinfo{series}{Koli Calling '14})}.
  \bibinfo{publisher}{Association for Computing Machinery},
  \bibinfo{address}{New York, NY, USA}, \bibinfo{pages}{91–97}.
\newblock
\showISBNx{9781450330657}
\urldef\tempurl%
\url{https://doi.org/10.1145/2674683.2674701}
\showDOI{\tempurl}


\bibitem[\protect\citeauthoryear{Weeda, Izu, Kallia, and Barendsen}{Weeda
  et~al\mbox{.}}{2020}]%
        {weeda2020towards}
\bibfield{author}{\bibinfo{person}{Renske Weeda}, \bibinfo{person}{Cruz Izu},
  \bibinfo{person}{Maria Kallia}, {and} \bibinfo{person}{Erik Barendsen}.}
  \bibinfo{year}{2020}\natexlab{}.
\newblock \showarticletitle{Towards an Assessment Rubric for EiPE Tasks in
  Secondary Education: Identifying Quality Indicators and Descriptors}. In
  \bibinfo{booktitle}{\emph{Koli Calling '20: Proceedings of the 20th Koli
  Calling International Conference on Computing Education Research}} (Koli,
  Finland) \emph{(\bibinfo{series}{Koli Calling '20})}.
  \bibinfo{publisher}{Association for Computing Machinery},
  \bibinfo{address}{New York, NY, USA}, Article \bibinfo{articleno}{30},
  \bibinfo{numpages}{10}~pages.
\newblock
\showISBNx{9781450389211}
\urldef\tempurl%
\url{https://doi.org/10.1145/3428029.3428031}
\showDOI{\tempurl}


\end{thebibliography}

\end{document}